\documentclass{elsart}
\journal{Physics Letters B}

\begin{document}

\newcommand{\Qed}{\rule{2.5mm}{3mm}}
\newcommand{\balpha}{\mbox{\boldmath {$\alpha$}}}

\begin{frontmatter}
\title{Massless particles of any spin obey linear equations of motion}
\author[BG]{Bojan Gornik},
\author[NMB]{ Norma Manko\v c Bor\v stnik}

\address[BG]{Faculty of Mathematics and Physics, University of
Ljubljana, Jadranska 19, Ljubljana 1000, Slovenia}
\address[NMB]{Primorska Institute for Natural Sciences and Technology,
C. Mare\v zganskega upora 2, Koper 6000, Slovenia}

\begin{abstract}
The proof is presented that the Poincar\' e symmetry determines the equations 
of motion for massless particles of any spin
in $2n$-dimensional spaces, which are linear in the momentum\footnote[1]
{After this paper appeared on hep-th W. Siegel let us know that he proposed\cite{wsp,wsb} the 
equations $
(S^{ab}p_b + w p^a=0)|\Phi\rangle $ which 
are linear in $p^a$-momentum, as well as in $S^{ab}$ for  
 all irreducible representations of massless fields in any $d$.
 Following derivations of this paper 
one easily proves that solutions of Siegel's equations belong to irreducible representations
of the Poincar\' e group. The proof is much simpler than for  our equations 
$(W^a=\alpha p^a)|\Phi\rangle$. Both equations are of course equivalent. Following our derivations
one finds that the constant $w$ in  the Siegel's equation is $w=l_n$, with $l_n$ defined 
 in this paper.
One derives our equation from  the Siegel's one 
for even $d$ after some  rather tedious calculations if multiplying it by 
$\varepsilon_{aca_1a_2 \ldots a_{d-3} a_{d-2}} S^{a_1 a_2} \cdots
S^{a_{d-3}a_{d-2}}$. }: 
$(W^a=\alpha p^a)|\Phi\rangle$,
with $W^a$ the generalized Pauli-Ljubanski vector.
\end{abstract}
\end{frontmatter}

\section{Introduction}

All theories with $d > 4$ have to answer the question  why Nature has made a choice of  four-dimensional
subspace with one time and three space coordinates and with the particular choice of charges beside the
spin degree of freedom for either fermions or bosons. One of us and Nielsen \cite{mankocnielsen00,mankocnielsen01}
 has 
prooved that in $d$-dimensional spaces, with even $d$, the spin degrees of freedom require $q$ time and $(d-q)$ space
dimensions, with $q$ which has to be odd. Accordingly in four-dimensional space Nature could only make a choice of 
the Minkowsky metric. This proof was made under the assumption that equations of motion are for massless
fields of any spin linear in the $d$-momentum $p^a,\; a = 0,1,2,3,5,\ldots,d$.
(In addition, also the Hermiticity of the equation of motion operator
as well as that this operator operates within an irreducible representation of the Lorentz group was 
required.)
Our experiences tell us that equations of motion of all known massless fields are linear in the four-momentum $p^a,
\; a = 0,1,2,3.$ We are refering to the Dirac equation of motion for massless spinor fields and the Maxwell or
Maxwell-like equations of motion for massless bosonic fields. One of us together with 
A. Bor\v stnik \cite{mankoc92,mankoc93,mankoc99,mankocborstnik98}  
has shown that 
the Weyl-like equations exist not only for fermions but
also for bosons.
For four
dimensional space-time Wigner \cite{wigner}  classified the representations of
the Poincar\' e group, connecting them with particles. The classification  of representations can also be found in 
Weinberg \cite{weinberg2}, for example.
According to these classifications,  equations of motion follow, when 
 constraining  a solution space to a certain Poincar\' e group representation.  
For spinors this leads to the Dirac equation and for vectors to the Maxwell equations \cite{fonda}. 

The aim of this paper is to briefly present the proof 
(the detailed version is presented in Ref. \cite{gornikmankoc})
that in even dimensional  spaces, for any $d=2n$, free massless fields $|\Phi\rangle$ (
$(p^a p_a = 0)\; |\Phi\rangle,\quad a= 0,1,2,3,5,\ldots,2n$)
of any spin satisfy
equations of motion which are  linear in the $d$-momentum $p^a = (p^0, \overrightarrow{p})$
\begin{eqnarray}
(W^a &=& \alpha p^a)\;|\Phi\rangle, \quad a = 0,1,2,3,5,\dots,d,
\nonumber\\
\quad  {\rm with } \quad \alpha &=& - \rho 2^{n-1} (n-1)! (l_{n-1}+n-2)\ldots (l_2 + 1) l_1
\label{meq}
\end{eqnarray}
and guarantee the validity of the equation $(p^a p_a = 0)\; |\Phi\rangle$.

  In Eq. (\ref{meq}) the operators $S^{ab}$ are the generators of the
Lorentz group $SO(1, d-1)$ in internal space, which is the space
of spin degrees of freedom.
Parameters $l_1, \ldots, l_n$ are eigenvalues of the operators of the Cartan (Eq.(\ref{cartan})) 
subalgebra of the 
algebra of the group $SO(1, d-1)$ 
on the maximal weight state (\ref{domweight}) of an irreducible representation.
Vector $W^a$ is the generalized Pauli-Ljubanski \cite{mankoc93} $d$-vector
\begin{eqnarray}
  W^a := \rho \,\varepsilon^{ab}{ }_{a_1 a_2\ldots a_{d-3}a_{d-2}} p_b S^{a_1 a_2}\ldots
    S^{a_{d-3}a_{d-2}}.
  \label{pauliljub}
\end{eqnarray}
The value of $\rho$ is irrelevant in the equations of motion (\ref{meq}) since it cancels out. 
It becomes relevant with
the introduction of $d-1$ vector \cite{mankocnielsen00,mankocnielsen01} 
\begin{equation}
S^i := \rho  \,\varepsilon^{0i}{ }_{a_1a_2 \ldots a_{d-3} a_{d-2}}
 S^{a_1a_2} \ldots S^{a_{d-3}a_{d-2}}.
\label{svector}
\end{equation}
We choose the value of $\rho$ in such a way that $S^i$ has eigenvalues independent of dimension $d$ and with values
familiar from four dimensions. For spinors, which are determined by Eqs.(\ref{spinor}) or equivalently by
$1/2 = l_n = \ldots = l_2 = \pm l_1$ we have
\begin{equation}
  \alpha = \pm \frac{1}{2},\quad \rho = \frac{2^{n-2}}{(2n-2)!}
\nonumber
\end{equation}
and for vector fields with $1 = l_n = \ldots = l_2 = \pm l_1$ we have
\begin{equation}
  \alpha = \pm 1,\quad \rho = \frac{1}{2^{n-1}(n-1)!^2}.
\nonumber
\end{equation}

The proof is made only for fields with no gauge 
symmetry and with a nonzero value of the {\it handedness} operator \cite{mankoc93,mankoc99}
$\Gamma^{(int)}$
\begin{eqnarray}
  \Gamma^{(int)}: &=& \beta \,\varepsilon_{a_1 a_2 \ldots a_{d-1}a_d} S^{a_1 a_2} \ldots
   S^{a_{d-1}a_d},
   \nonumber\\
  \beta &=& \frac{i}{2^n n! (l_n+n-1)\ldots(l_2+1)l_1}, \quad d=2n, 
  \label{gamaint}
\end{eqnarray}
which commutes with all the generators of the Poincar\' e group.
We choose $\beta$  so that $\Gamma^{(int)} = \pm 1$
on representations with nonzero handedness.
For spinors (Eq.(\ref{spinor}))  $\beta=(2^n i)/(2n)!$,
while for  vector fields it is $\beta =i/(2^n (n!)^2)$.

We prove that for spinors in $d$-dimensional space Eq.(\ref{meq}) is equivalent to the equation
\begin{equation}
(\Gamma^{(int)} p^0 = \frac{1}{|\alpha|} \;\overrightarrow{S}\cdot \overrightarrow{p})|\Phi\rangle.
\label{meq1}
\end{equation}
with $1/|\alpha|$ equal to $2$ for any $d=2n$ while for a general spin Eq.(\ref{meq})  may
impose additional conditions on the field. 

We recognize the generators $S^{ab}$ to be of the spinorial character, if they fulfil the relation
\begin{equation}
\{S^{ab}, S^{ac}\} = \frac{1}{2} \; \eta^{aa}\; \eta^{bc}, \;\;{\rm no\;\; summation\;\; over\;\; a},
\label{spinor}
\end{equation}
with $\{A,B\} = AB + BA $.

In this paper the metric is, independantly of the dimension,  assumed to be the Minkowsky metric with 
$\eta^{ab}= \delta^{ab} (-1)^A, \;A= 0\; {\rm for}\; a=0 \;{\rm and}\; A=1$, otherwise.

We present  the proof in steps introducing only the very needed quantities and assuming that the reader 
can find the rest  in text-books, as well as in 
Ref. \cite{gornikmankoc}.

\section{Irreducible representations of the Lorentz group} 

We denote an irreducible representation of the Lorentz group $SO(1, d-1)$ by
the weight of the dominant weight state of the representation \cite{giorgi}. 

The Lie algebra of $SO(1, d-1)$ is spanned by the generators $S^{ab}$, which satisfy the commutation
relations $
[S^{ab}, S^{cd}] = 
  i(\eta^{ad} \; S^{bc} + \eta^{bc} \; S^{ad} - \eta^{ac} \; S^{bd} - 
\eta^{bd}\; S^{ac}).$ We choose  the $n$ commuting operators  of the Lorentz group $SO(1,d-1)$ as follows
  \begin{equation}
-i S^{0 d}, S^{1 2}, S^{3 5}, \ldots, S^{d-2 \; d-1}
\label{cartan}
\end{equation}
and call them
${C}_0, { C}_1, { C}_2, \ldots, { C}_{n-1}\;$  respectively. 
We say that a state $|\Phi_w\rangle$ has the weight
$(w_0, w_1, w_2, \ldots, w_{n-1})$ if the following equations hold
\begin{equation}
 { C}_j|\Phi_w\rangle = w_j|\Phi_w\rangle,\quad j=0,1,\ldots,n-1. 
\label{weight}
\end{equation}
According to the definition of the operators (Eq.(\ref{cartan})), weight components
$w_0, w_1, w_2, \ldots, w_{n-1}$ are always real numbers.

  We introduce in a standard way \cite{matematika} the raising and the lowering operators
\begin{equation}
 E_{j k}(\lambda, \mu) := \frac{1}{2}((-i)^{\delta_{j 0}} S^{j_- k_-} + i\lambda S^{j_+ k_-}
 - i^{1+\delta_{j 0}}\mu S^{j_- k_+} - \lambda \mu S^{j_+ k_+} ), 
\label{raislow}
\end{equation}
with $0\le j<k\le n-1,\,\, \lambda,\mu=\pm 1\;$ and $\;0_- = 0, \;0_+ = d,\; 1_- = 1,\; 1_+ = 2,
\; 2_- = 3, \; 2_+ = 5\; $ and so on. Due to the commutation relations 
\begin{equation}
[ E_{j k}(\lambda, \mu), { C}_l ] = (\delta_{j l}\lambda + \delta_{k l}\mu)  E_{j k}(\lambda, \mu), 
\nonumber
\end{equation} 
if the state 
$|\Phi_w\rangle$ has the weight $(w_0, w_1, \ldots, w_{n-1})$ then the state 
$E_{j k}(\lambda, \mu) |\Phi_w\rangle$ has the weight $(\ldots, w_j + \lambda, \ldots, w_k + \mu, \ldots)$.
We call the state $|\Phi_{l}\rangle$ with the property
\begin{eqnarray}
  E_{j k}(+1, \pm 1)|\Phi_{l}\rangle = 0, \quad 0\le j < k \le n - 1,
\label{domweight}
\end{eqnarray}
 the dominant weight state. All the other states of an irreducible representation  are
obtained by the application of the generators $S^{ab}$. We shall denote an irreducible 
representation of the Lorentz group
$SO(1, d-1)$ by the weight of the dominant weight state: 
$(l_n, l_{n-1}, \ldots, l_2, l_1)$.
Numbers $l_n, l_{n-1}, \ldots, l_2, l_1$ are either all integer or all half integer and 
satisfy $
  l_n\ge l_{n-1}\ge \ldots \ge l_2\ge |l_1|.$

We can write Eq.(\ref{domweight}) in an equivalent way
\begin{eqnarray}
(S^{0 i} + S^{i d}) |\Phi_{l}\rangle = 0, \quad i = 1, 2, 3, 5, \ldots, d - 1  
\nonumber\\
(S^{1 i} + i S^{2 i}) |\Phi_{l}\rangle = 0, \quad i = 3, 5, \ldots, d - 1,
\nonumber\\
(S^{3 i} + i S^{5 i}) |\Phi_{l}\rangle = 0, \quad i = 6, 7 \ldots, d - 1,\,\,\,{\rm and\,\,so\,\,on.}
\label{dom}
\end{eqnarray}

By applying  $\Gamma^{(int)} = \beta\varepsilon_{a_1 a_2\ldots a_{d-1} a_d} S^{a_1 a_2} S^{a_3 a_4} \ldots  $ (Eq.(\ref{gamaint}))
on a state with the dominant weight and taking into account Eqs.(\ref{dom}), we find
\begin{equation}
 ( \Gamma^{(int)}   = 2^n n! i \beta  (l_n + n - 1)(l_{n-1} + n - 2)\ldots(l_2 + 1) l_1)|\Phi_l\rangle. 
\label{gama1}
\end{equation}
In order to obtain $\Gamma^{(int)}= \pm 1$  for any irreducible representation in any $d=2n$, $\beta$ 
must be the one, presented in Eq.(\ref{gamaint}).

\section{The unitary discrete massless representations of the Poincar\' e group}

  The generators of the Poincar\' e group, that is the generators of translations $p^a$ and 
the generators of the Lorentz transformations $M^{ab}$ (which form the Lorentz group),
fulfil in any dimension $d$, even or odd, the commutation relations:
\begin{eqnarray}
[p^a, p^b] &=& 0,
\nonumber
\\
\;[M^{ab}, M^{cd}]
&=&  i(\eta^{ad} \; M^{bc} + \eta^{bc} \; M^{ad} - \eta^{ac} \; M^{bd} - 
\eta^{bd}\; M^{ac}),
\nonumber
\\
\;[M^{ab}, p^c]
&=& i (\eta^{bc}\; p^a - \eta^{ac}\; p^b). \nonumber
\end{eqnarray}
  For momenta $p^a$ appearing in an irreducible massless representations of the Poincar\' e group 
it holds $p^a p_a = 0,\quad  p^0 > 0 \,\,{\rm or}\,\, p^0 < 0$
(we omit the trivial case $p^0=0$). We denote $r=p^0/|p^0|$. 
The Poincar\' e group representations are then characterized by the representation of the 
{\it little group}, which is a subgroup of the Lorentz group leaving
some fixed $d$-momentum $p^a={\bf k}^a$, satisfying equation $ {\bf k}^a {\bf k}_a =0$, unchanged.
Making the choice of ${\bf k}^a = (r k^0,0,\ldots,0,k^0)$, with $ r =\pm 1, k^0 > 0$, 
the infinitesimal generators of the little group 
$\omega_{bc} M^{bc},\,(\omega_{bc}=-\omega_{cb})$ can be found by requiring that, 
when operating on the $d$-vector ${\bf k}^a$
give zero, so that accordingly the corresponding group transformations leave the $d$-vector ${\bf k}^a$ 
unchanged: 
$(\omega_{bc} M^{bc}){\bf k}^a = 0$.
It is easily checked that this requirement leads to equations
$\omega_{0d}= 0,
\;\;\omega_{i0}+r\omega_{id}=0,\quad i = 1,2,3,5,\ldots d-1.$

All $\omega_{bc} M^{bc}$ with $\omega_{bc}$ subject to conditions  $(\omega_{bc} M^{bc}){\bf k}^a = 0$ form the
Lie algebra of the little group. We choose the following basis of the little group
Lie algebra
\begin{eqnarray}
\Pi_i = M^{0i} + r M^{id},\quad i = 1,2,3,5,\ldots,d-1\quad 
\nonumber
\\
{\rm and \; all\;\;} M^{ij}, \quad i, j = 1,2,3,5,\ldots,d-1.
\label{glg1}
\end{eqnarray}
One finds
\begin{eqnarray}
[\Pi_i, \Pi_j] = 0, \quad [\Pi_i, M^{jk}] = i (\eta^{ij} \Pi_k - \eta^{ik} \Pi_j).
\label{glgc}
\end{eqnarray}
We are interested only in unitary discrete representations of the Poincar\' e group. This means that
the states in the representation space can be labeled by the momentum and an additional
label for internal degrees of freedom, which can only have {\it discrete} values.

{\it Lemma 1:} For a discrete representation of the Poincar\' e group
operators $\Pi_i$ give zero $\Pi_i |\Phi_a\rangle = 0.$

{\it Proof:} We may arrange the representation space of the little group so that the commuting operators
$\Pi_1, \ldots, \Pi_{d-1}$ are diagonal:
$\Pi_i |\Phi_a\rangle = b^i_a |\Phi_a\rangle$. Making the rotation 
$e^{i\theta M^{ij}}|\Phi_a\rangle$ for a continious set of values of the parameter $\theta$ and for the chosen
indices $i,j$ we find that the states $e^{i\theta M^{ij}}|\Phi_a\rangle$ can only have the discrete
eigenvalues for $\Pi_i$ if $b^i_a = 0$ for all $a$ and $i$. 
(The detailed version of the proof can be found in Ref. \cite{gornikmankoc}).
\hfill\Qed

We now introduce the decomposition of the generators of Lorentz transformations to external and internal space.
We write $M^{ab} = L^{ab} + S^{ab}$ where $L^{ab} = x^a p^b - x^b p^a$ and $S^{ab}$ are
the generators of the Lorentz transformations in internal space (ie. spin generators).
We see that on the representation space of the little group with the choice 
${\bf k}^a =(r k^0, 0, \ldots, 0, k^0)$, on which $(L^{0i} + r L^{id})|\Phi_a\rangle= 
0,$ the following holds:
\begin{equation}
\Pi_i |\Phi_a\rangle = \Pi^{(int)}_i |\Phi_a\rangle = 0, \quad \Pi^{(int)}_i := (S^{0i} + r S^{id}), \quad 
{\rm for \;\; each \;\;} i.
\label{piint}
\end{equation}
The only little group generators, which are not necesserily zero on the 
representation space, are $M^{ij},
i, j = 1, 2, \ldots, d-1$ and they form the Lie algebra of $SO(d-2)$.
Since the irreducible discrete representations of the Poincar\' e group in 
$d(=2n)$ dimensions for massless particles are determined by the irreducible representations
of the group $SO(d-2)$ we will  denote the former with the same symbol as the 
latter with an additional label $r$ (ie. energy sign):
$(l_{n-1}, l_{n-2}, \ldots, l_2, l_1; r)$.

{\it Lemma 2:} On the representation space of an irreducible
massless representation of the Poincar\' e group $(l_{n-1},\ldots,l_1; r)$
the equation (\ref{meq}) holds with $\alpha = -\rho r 2^{n-1} (n-1)! (l_{n-1}+n-2)\ldots(l_2+1)l_1$.

{\it Proof:} First, we prove the lemma on the representation space of the little group
for the choice $p^a = {\bf k}^a = (r k^0, 0,\ldots,0,k^0), \; k^0 > 0, r=\pm 1$.
We begin with the cases $a=1,2,\ldots,d-1$ in Eq.(\ref{meq})
\begin{eqnarray}
  W^a/\rho|\Phi_a\rangle = (\varepsilon^{a a_1}{ }_{a_2 a_3\ldots a_{d-1}} p_{a_1} M^{a_2 a_3} M^{a_4 a_5}\ldots)|\Phi_a\rangle 
 = \nonumber\\
   -2 k^0 (n-1) {\balpha}^{ai} \Pi_i|\Phi_a\rangle,
  \label{wijenic}
\end{eqnarray}
where
\begin{equation}
{\balpha}^{ai} := 
  \varepsilon^{0dai}{ }_{a_1 a_2 \ldots a_{d-4}} M^{a_1a_2} \ldots M^{a_{d-5}a_{d-4}},\quad
[\Pi^i, {\balpha}^{ai}] = 0.
\label{balfa}
\end{equation}
  Taking into account Eq.(\ref{piint}) we  conclude that for $a=1,2,\ldots,d-1$, $\;W^a |\Phi_a\rangle = 
p^a |\Phi_a\rangle =0$.

For  $a=0$ and $a=d$ we obtain
\begin{eqnarray}
  W^0/\rho|\Phi_a\rangle = 
  (-r) p^0 \Gamma^{(int)}_{d-2}/\beta|\Phi_a\rangle,
  \nonumber \\
  W^d/\rho|\Phi_a\rangle =  
  (-r) p^d \Gamma^{(int)}_{d-2}/\beta|\Phi_a\rangle,
  \label{wininic}
\end{eqnarray}
where $\Gamma^{(int)}_{d-2}$ is the handedness operator corresponding to the subgroup $SO(d-2)\le SO(1, d-1)$ acting
on coordinates $1, \ldots, d-1$. Substituting for $\Gamma^{(int)}_{d-2}$ the appropriate value (Eq.(\ref{gama1}) with
$n-1$ instead of $n$ and without the $i$ factor) we conclude the proof for
the choice $p^a = {\bf k}^a = (r k^0,0,\ldots,0,k^0)$.
To extend the proof for  Eq.(\ref{meq}) to the whole representation space one only has to note that 
Eq.(\ref{meq}) is in a covariant form and must therefore hold generally.
\hfill \Qed

  Since Eq.(\ref{meq}) with $\alpha$ from Eq.(\ref{meq}) 
  holds on the Poincar\' e group representation
$(l_{n-1},\ldots,l_1; +1)$ it is  a candidate for an equation of motion. 

\section{Equations of motion for free massless fields of any spin in $d=2n$}

What we have to prove is that the solutions of  Eq.(\ref{meq}) belong to an irreducible
unitary discrete representation, possibly with a degeneracy in the energy sign.

In all proofs that follow we shall make the choice $p^a =  {\bf k}^a = 
(r k^0,0, \ldots,0,k^0)$, $k^0 > 0, r = \pm 1,$ since then 
the covariance of Eq.(\ref{meq}) guarantees that the proofs are valid for general $p^a$.

{\it Lemma 3:} On the space with internal Lorentz group $SO(1, d-1)$ 
representation $(l_n, l_{n-1}, \ldots, l_2, l_1),$ states satisfying 
the discreteness condition of Eq.(\ref{piint}) for the little group, form the 
$SO(d-2)$-irreducible representation space $(l_{n-1}, \ldots, r l_1)$, where the subgroup $SO(d-2)\le
SO(1,d-1)$ acts on coordinates $1, 2, \ldots, d-1$.

{\it Proof:} We present an outline of the proof for $r=1$ only. The case $r=-1$ is treated similarly 
(see \cite{gornikmankoc}).

  Eq.(\ref{glgc}) implies that the space of solutions 
$\{|\Phi_a\rangle\}$ of Eq.(\ref{piint}) forms an $SO(d-2)$-invariant space, where 
$SO(d-2)\le SO(1, d-1)$ acts on coordinates $1, 2, \ldots, d-1$. 
To show that $\{|\Phi_a\rangle\}$ is irreducible and corresponds to the 
$SO(d-2)$-representation
$(l_{n-1}, \ldots,  l_1)$, we choose any 
state $|\Phi_a\rangle\in\{|\Phi_a\rangle\}$ with a $SO(d-2)$-dominant weight
and prove both that it is unique
up to a scalar multiple and has $SO(d-2)$-weight $(l_{n-1}, \ldots,  l_1)$. 
  The set of states 
$\{|\Phi_a\rangle\}$ is  nontrivial, because the $SO(1, d-1)$-dominant weight state is
in $\{|\Phi_a\rangle\}$.

Since $|\Phi_a\rangle$ has a $SO(d-2)$-dominant weight the last two lines in Eqs.(\ref{dom}) must hold.
 Eq.(\ref{piint}) then implies that the first line of Eq.(\ref{dom}) must also hold.
It follows then that  the state $|\Phi_a\rangle$ has a $SO(1, d-1)$-dominant weight 
and is (up to a scalar multiple) unique and the  proof that the corresponding 
representation of the Poincar\' e group is determined by all but the first 
dominant weight component
of the internal Lorentz group representation $(l_{n-1}, l_{n-2}, \ldots, l_2, l_1)$ is complete.
\hfill\Qed


It remains to answer the question: when are the solutions of  equations
$\; (W^a = \alpha p^a)|\Phi\rangle,$ with $ \alpha = - \rho 2^{n-1}(n-1)!
(l_{n-1}+(n-2))\ldots (l_2 + 1) l_1$
on the space with internal Lorentz group $SO(1, d-1)$ 
representation $(l_n, l_{n-1}, \ldots, l_2, l_1),$ 
exactly those described by the lemma 3, ie. $(l_{n-1},\ldots, r l_1; r), r = \pm 1$.

{\it Lemma 4:} On the space with internal Lorentz group $SO(1, d-1)$
representation $(l_n, l_{n-1}, \ldots, l_1),$ where $
  l_1 \not= 0  \Longleftrightarrow \Gamma^{(int)} \not= 0,$
the solutions of Eq.(\ref{meq}) are exactly $(l_{n-1},\ldots,rl_1; r),$ with $r = \pm 1$.

{\em Proof:} 
First, we take  the simplest case
of $d=4$, proving that on the space with internal Lorentz group $SO(1, 3)$ 
representation $(l_2, l_1)$ the solutions of  Eq.(\ref{meq}) are exactly
$(l_1; +1)$ and $(-l_1; -1)$.

In this case, the first equation of Eqs.(\ref{meq}), with $a=1,2$ reads 
$(\Pi^{(int)}_2 = 0)|\Phi\rangle,\;\;\; (\Pi^{(int)}_1 = 0)|\Phi\rangle,$
which are exactly  Eq.(\ref{piint}) and  lemma 3 completes 
the proof since cases $a=0,3$ both give the equation $S^{12}|\Phi\rangle = r l_1 |\Phi\rangle$
which imposes no additional constraints on representation spaces $(l_1; +1)$ and $(-l_1; -1)$.

  We  can proceed now with the general case $d\ge 6$.
>From Eqs.(\ref{wijenic}), (\ref{wininic}) we know that Eq.(\ref{meq}) can be written for our choice of $p^a$
as follows
\begin{eqnarray}
{\balpha}^{ij} \Pi^{(int)}_j |\Phi\rangle &=& 0,\quad {\rm for}\;\;a=i,
\nonumber\\
 -\rho \varepsilon^{0d}{ }_{a_1 a_2\ldots a_{d-3}a_{d-2}} S^{a_1 a_2} S^{a_3 a_4}\ldots|\Phi\rangle &=& 
  \alpha|\Phi\rangle, \quad {\rm for} 
 \;\; a=0 \;\; {\rm  and} \;\; a=d,
 \label{alphaijpi}
 \end{eqnarray}

with $ {\balpha}^{ij}$ defined in Eq.(\ref{balfa}).

By lemma 3 it is sufficient to show that every solution $|\Phi\rangle$ of Eqs.(\ref{alphaijpi})
 satisfies the condition
\begin{eqnarray}
\Pi^{(int)}_i|\Phi\rangle = 0, \quad {\rm for}\,\, i=1,2,\ldots,d-1.
\label{piintphi}
\end{eqnarray}

According to the proof of lemma 3 the 
dominant weight of the  group $SO(1,d-1)$
satisfies all the above equations  and   we  can conclude that the space ${ V}$
of solutions of equations in Eq.(\ref{alphaijpi}) is nontrivial. 
It is also $SO(d-2)$-invariant due to the commutation relations
$ [{\balpha}^{ij} \Pi^{(int)}_j, S^{kl}] = i(\eta^{ik}{\balpha}^{lj} \Pi^{(int)}_j - \eta^{il}
{\balpha}^{kj} \Pi^{(int)}_j)$ and $
[\varepsilon^{0d}{ }_{a_1a_2 \ldots a_{d-3}a_{d-2}} S^{a_1a_2} \ldots S^{a_{d-3}a_{d-2}},
S^{kl}] = 0$ for $i,j,k,l=1,2,\ldots,d-1$ .

To prove that Eq.(\ref{piintphi}) holds on $ V$
it suffices to prove that Eq.(\ref{piintphi}) holds on any $SO(d-2)$-dominant weight state
in $ V$. This is done with the aid of Eqs.(\ref{alphaijpi}), (\ref{dom}) and some lengthy but
elementary calculations (details can be found in \cite{gornikmankoc}).

\hfill \Qed

We may now write down the main result of this letter.

\vspace{2mm}

{\it On the space with internal Lorentz group
representation $(l_n, l_{n-1}, \ldots, l_1)$ where $l_1\not=0$
equations
\begin{equation}
  (W^a = \alpha p^a)|\Phi\rangle, \alpha = -\rho 2^{n-1}(n-1)!(l_{n-1}+n-2)\ldots(l_2+1)l_1
  \label{glavna_en2}
\end{equation}
are equations of motion for massless particles corresponding to the following representations of the
Poincar\' e group}
\begin{equation}
  (l_{n-1}, \ldots, l_1; +1)\quad{\rm and}\quad (l_{n-1},\ldots,-l_1;-1),
\end{equation}
{\it where the masslessness condition $(p_a p^a=0)|\Phi\rangle$ is not needed, since it follows from (\ref{glavna_en2})
and they are linear in the $p^a$-momentum.

With the aid of Eqs. (\ref{gama1}), (\ref{gamaint}) the equation of motion can also be written as
\begin{eqnarray}
  (W^a = |\alpha| \Gamma^{(int)} p^a)|\Phi\rangle, \nonumber\\
  |\alpha| = \rho 2^{n-1}(n-1)!(l_{n-1}+n-2)\ldots(l_2+1)|l_1|.
  \label{glavna_en3}
\end{eqnarray}
This equation is convenient when dealing with positive and negative handedness on the same footing
(an example of this is the Dirac equation) since $|\alpha|$ is independent of the
sign of $l_1$.}

We note that the particular value of $\rho$ is irrelevant in Eqs. (\ref{glavna_en2}), (\ref{glavna_en3}) since $\rho$
is found in both the lefthand and the righthand side of  equations and thus cancels out. (The value of $\rho$ becomes
relevant, as we have already said, when dealing with the particular spin when it is used to insure that
the operators $S^i$  have the familiar values independent of the dimension.)

Making a choice of $a=0$ one finds
\begin{equation}
(\overrightarrow{S}.\overrightarrow{p}= |\alpha| \Gamma^{(int)} p^0)|\Phi\rangle.
\label{sag}
\end{equation}
Since $(\Gamma^{(int)})^2=1$, one immediately finds that 
$ |\alpha| = |\overrightarrow{S}.\overrightarrow{p}|/|p^0|$.
The rest of equations make no additional requirements for spinors, while this is not the case for other 
spins.

\section{Concluding remarks}

We have proven  in this letter that massless fields of any spin (with  nonzero handedness
and no gauge symmetry) in $d=2n$-dimensional 
spaces, if having the Poincar\' e symmetry, obey the {\em linear equations
of motion }(Eq.(\ref{meq})).  We have limited our proof to only even $d$, because the operator 
for  handedness 
Eq.(\ref{gamaint}), needed in the proof, as well as the $d$-vector (Eq.(\ref{svector})),
can only be defined in even-dimensional spaces. (The generalization of the proof to all $d$ and any signature
is under consideration.)

We know that in four-dimensional space the Weyl equation is linear in the four-momentum
and so are the Maxwell equations \cite{mankocborstnik98},  describing massless spinors and  
massless  vectors, respectively. In $d$-dimensional spaces the operator of handedness for spinors can
be defined not only for even but also for odd dimensions. Also the Dirac-like equations 
exist  for any dimension \cite{mankoc99,gornikmankoc} and follow for even $d$  from 
Eq.(\ref{meq}) if we take into account that for spinors $S^{ab} = -i[\gamma^a, \gamma^b]/4$:
\begin{eqnarray}
(\gamma^a p_a = 0)|\Phi\rangle,
\label{dirac}
\end{eqnarray}
with $\gamma^a$ matrices defined for $d$-dimensional spaces \cite{mankoc99,gornikmankoc,giorgi}.

Similarly it follows from Eq.(\ref{meq}) \cite{gornikmankoc} that  vectors in $d$-dimensional space obey the equations
of motion
\begin{equation}
  p_a F^{a a_1\ldots a_{n-1}} = p_a \varepsilon^{a a_1\ldots a_{n-1}}{}_{b_1 b_2\ldots b_n} F^{b_1 b_2\ldots b_n} = 0
\label{vectorem}
\end{equation}
where $F^{abcd..}$ is a totally antisymmmetric tensor field.

\section{Acknowledgement } 

This work was supported by Ministry of
Science and Technology of Slovenia. One of us (N.M.B.) wants to thank for  many fruitful
discussions to Holger Bech Nielsen and Anamarija Bor\v stnik.

\end{document}